\documentclass{article}

\usepackage[T1]{fontenc}

\usepackage{PRIMEarxiv}

\usepackage[table,dvipsnames]{xcolor}
\usepackage{amsmath,amsfonts}
\usepackage[ruled,vlined]{algorithm2e}
\usepackage{graphicx}
\usepackage{mathtools}
\usepackage{mathrsfs}
\usepackage{caption}
\usepackage{amsthm}
\usepackage{amssymb}
\usepackage[colorlinks=true]{hyperref}
\hypersetup{
  linkcolor=black,
  citecolor=black,
  urlcolor=[RGB]{5,115,185}
}
\usepackage{microtype}
\usepackage{multirow}						
\usepackage{tabularx} 						
\usepackage{colortbl}
\usepackage{stmaryrd}
\usepackage{sidecap}
\usepackage{titlesec}
\usepackage{authblk}
\usepackage{listings}
\usepackage{booktabs} 
\usepackage{subcaption}

\usepackage[sorting=none, backend=biber]{biblatex} 
\definecolor{LightGray}{rgb}{0.8,0.8,0.8}
\definecolor{LightCyan}{rgb}{0.74,0.83,0.9}
\definecolor{DarkBlue}{rgb}{0,0.28,0.67}
\definecolor{blizzardblue}{rgb}{0.67, 0.9, 0.93}
\definecolor{inchworm}{rgb}{0.7, 0.93, 0.36}
\definecolor{coralred}{rgb}{1.0, 0.25, 0.25}
\definecolor{celadon}{rgb}{0.67, 0.88, 0.69}

\DisableLigatures{encoding = *, family = * }

\newcommand{\sherpa}{\href{https://www.sherpa.ai/}{\textcolor{DarkBlue}{Sherpa.ai} }}

\SetSymbolFont{stmry}{bold}{U}{stmry}{m}{n}

\definecolor{DarkColor}{gray}{0.75}			
\definecolor{LightColor}{gray}{0.9}
\definecolor{LightGrey}{rgb}{0.98,0.98,0.98}
\definecolor{DarkGrey}{rgb}{0.83,0.83,0.83}
\definecolor{BaseColor}{rgb}{0.10,0.10,0.20}
\definecolor{TextColor}{RGB}{58,88,119}
\definecolor{LightTextColor}{RGB}{229,233,205}
\definecolor{DarkTextColor}{RGB}{25,29,1}
\definecolor{NeutralBg}{rgb}{0.92,0.92,0.92}
\definecolor{LightYellow}{RGB}{255,255,102}
\definecolor{DarkOrange}{RGB}{255,90,0}
\definecolor{Green}{RGB}{0,128,0}
\definecolor{White}{gray}{1}

\usepackage{etoolbox}
\makeatletter
\patchcmd{\@begintheorem}{\textit}{\textbf}{}{}
\makeatother

\newcolumntype{L}{>{\raggedright\arraybackslash}X}   
\newcolumntype{Y}{>{\centering\arraybackslash}X}

\newcolumntype{Z}{>{\raggedright\arraybackslash}p{3.5cm}}
\newcolumntype{A}[1]{>{\hspace*{-#1}\centering\arraybackslash}X}
\newcolumntype{B}{>{\centering\arraybackslash}p{5cm}}
\renewcommand{\arraystretch}{1.3}  

\newtheorem{remark}{Remark}

\numberwithin{equation}{section}
\numberwithin{subsection}{section}

\usepackage{etoolbox}

\AtBeginEnvironment{tabular}{\small}

\title{Federated Cyber Defense: Privacy-Preserving Ransomware Detection Across Distributed Systems}

\author{%
  {\LARGE \href{https://sherpa.ai/}{Sherpa.ai}}\\
  research@sherpa.ai
}

\fancypagestyle{firstpagestyle}{%
  \fancyhf{}%
  \fancyhead[R]{\includegraphics[scale=1.0]{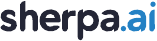}} 
  \fancyfoot[C]{\thepage}%
}

\begin{document}
\maketitle
\thispagestyle{firstpagestyle}

\begin{abstract}

Detecting malware, especially ransomware, is essential to securing today’s interconnected ecosystems, including cloud storage, enterprise file-sharing, and database services. Training high-performing artificial intelligence (AI) detectors requires diverse datasets, which are often distributed across multiple organizations, making centralization necessary. However, centralized learning is often impractical due to security, privacy regulations, data ownership issues, and legal barriers to cross-organizational sharing. Compounding this challenge, ransomware evolves rapidly, demanding models that are both robust and adaptable.

In this paper, we evaluate Federated Learning (FL) using the \sherpa FL platform, which enables multiple organizations to collaboratively train a ransomware detection model while keeping raw data local and secure. This paradigm is particularly relevant for cybersecurity companies (including both software and hardware vendors) that deploy ransomware detection or firewall systems across millions of endpoints. In such environments, data cannot be transferred outside the customer’s device due to strict security, privacy, or regulatory constraints. Although FL applies broadly to malware threats, we validate the approach using the Ransomware Storage Access Patterns (RanSAP) dataset.

Our experiments demonstrate that FL improves ransomware detection accuracy by a relative 9\% over server-local models and achieves performance comparable to centralized training. These results indicate that FL offers a scalable, high-performing, and privacy-preserving framework for proactive ransomware detection across organizational and regulatory boundaries.

\end{abstract}

\begin{figure}[h]
    \centering
    \includegraphics[width=0.9\linewidth]{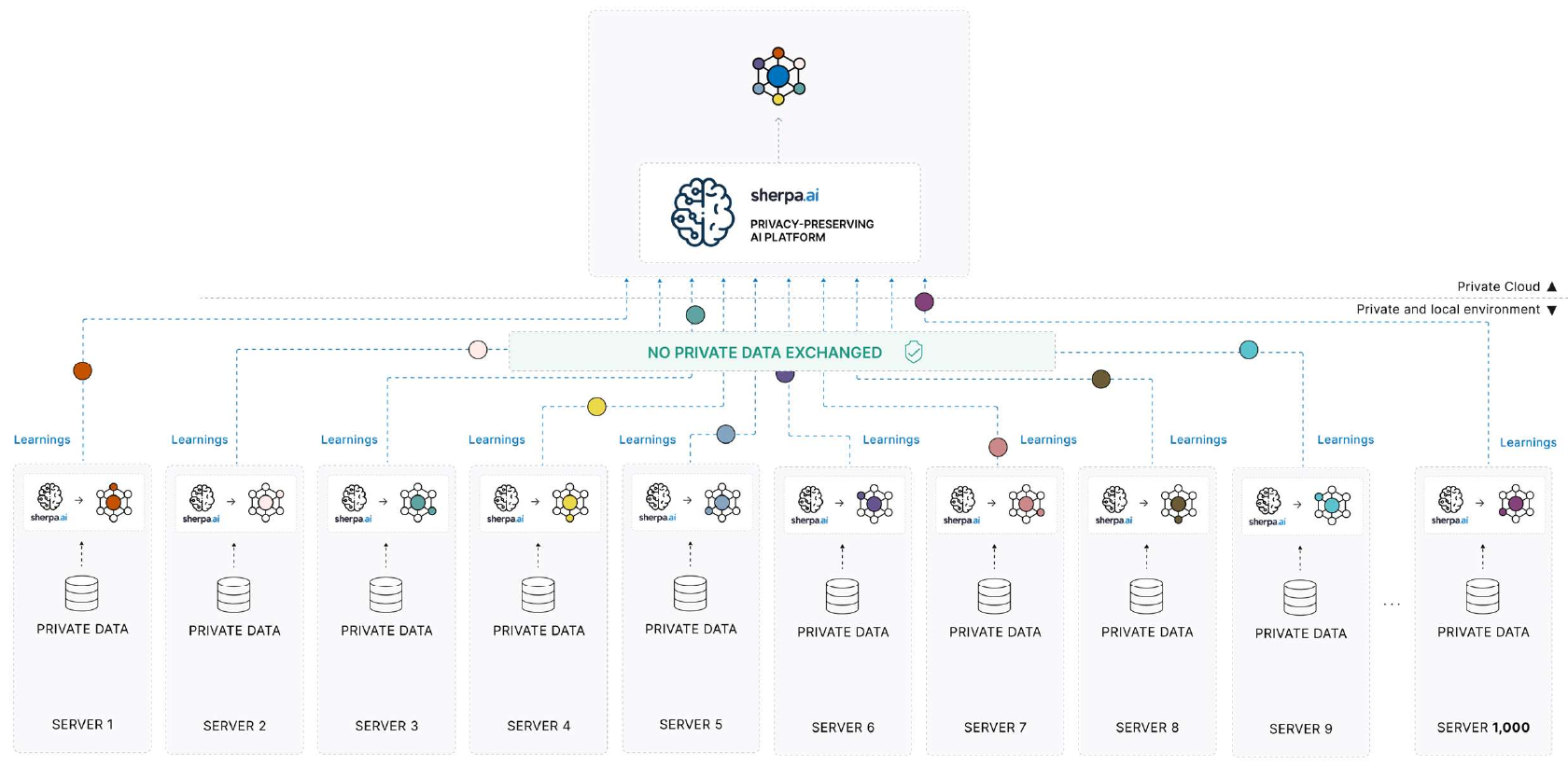}
    \caption{Privacy-preserving FL over customers' endpoints (servers): each server trains locally on its own logs (data) and shares only model updates; no raw data leaves the servers.}
    \label{fig:multiple_nodes}
\end{figure}

\section{Introduction}

Malware, short for malicious software, refers to any software intentionally designed to disrupt, damage, or gain unauthorized access to computer systems. Common categories include viruses, worms, Trojan horses, ransomware, and spyware~\cite{techtarget_definition}. Among these, ransomware has become especially disruptive, targeting both public and private infrastructure with attacks that encrypt data and demand ransom payments. The global impact of ransomware has escalated dramatically, with damages projected to exceed trillions of dollars annually.

Traditional malware detection approaches primarily rely on signature-based methods, which compare files against a database of known malware signatures. While effective for previously identified threats, these methods are inherently reactive and vulnerable to novel, obfuscated, and polymorphic variants~\cite{obfuscation_limitation}. To address these limitations, Machine Learning (ML) and behavior-based detection techniques have been proposed, focusing on patterns of activity rather than static features~\cite{aslan2020comprehensive}.

However, ML-based methods introduce new challenges, particularly in terms of \textit{data requirements}. These models require access to large-scale, diverse, and representative datasets to generalize effectively. In the case of ransomware detection, this includes telemetry data such as file system activity, process behavior, network communications, and cryptographic operations. Unfortunately, data sharing across organizations is often constrained by \textit{privacy laws, proprietary concerns,} and the risk of \textit{data leakage}. These barriers are particularly stringent in domains such as finance, healthcare, and industrial manufacturing, sectors that are frequent targets of ransomware and maintain strict confidentiality standards.

\textbf{Federated Learning} (FL)~\cite{mcmahan2017communication} has emerged as a compelling solution to these challenges by enabling collaborative model training across distributed nodes without exposing local data. In the FL paradigm, each participant trains a local model using its private data and shares only encrypted model parameters or updates with a central aggregator. This design preserves data privacy and aligns with regulatory frameworks~\cite{gdpr2016},~\cite{lopd2018},~\cite{hipaa1996},~\cite{CCPA2018}, etc. An overview of the deployment on customers' endpoints (servers) is shown in Figure~\ref{fig:multiple_nodes}.

In this paper, we explore the application of FL to malware detection, with a particular focus on ransomware detection under data privacy constraints. We analyze how FL frameworks, especially Horizontal FL (HFL)~\cite{liu2022towards}, can be employed to enhance detection robustness across heterogeneous environments while respecting organizational and legal boundaries on data sharing. We then compare our results to the two limit training scenarios: a \textit{centralized} one, with all available data (no privacy), and a \textit{local} node training using only its private dataset.

\section{Problem Formulation}

This section defines the use case explored in this paper: ransomware detection. We begin by introducing the concept of ransomware detection, followed by a formalization of the corresponding classification problem.

\subsection{Ransomware Detection}
Ransomware detection refers to the process of identifying malicious software that encrypts or locks access to a victim's data or systems, typically demanding a ransom payment for restoration. This task involves analyzing system behavior, network traffic, and file activity to uncover patterns that indicate the presence of ransomware, such as rapid file encryption, unauthorized file modifications, or anomalous process activity. An effective detection system aims to identify ransomware as early as possible—ideally before significant damage occurs— while minimizing false positives that could disrupt legitimate operations. In practice, the cost of failing to detect a true ransomware attack (false negative) is often far greater than the inconvenience of flagging benign activity (false positive). The primary objective is to ensure timely and accurate detection to protect critical data, maintain operational continuity, and prevent financial and reputational harm.

\begin{figure}[h]
    \centering
    \includegraphics[width=0.9\linewidth]{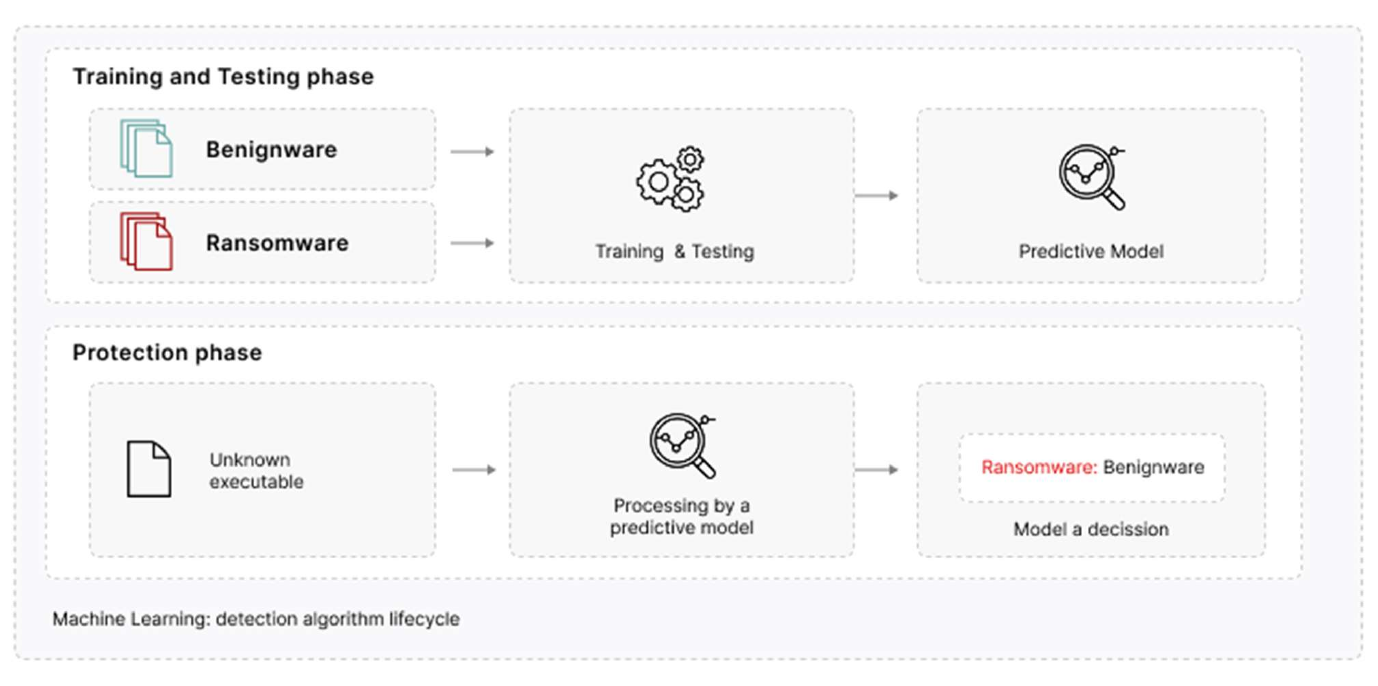}
    \caption{ML lifecycle for ransomware detection. During the training phase, labeled data comprising benign software (benignware) and known ransomware samples is used to train and validate a predictive model. In the protection phase, the trained model is deployed to classify unseen executables, enabling real-time ransomware detection based on their behavioral or static characteristics. Image modified from Herrera et al.~\cite{herrera2023dynamic}}
    \label{fig:detection-lifecycle}
\end{figure}

Figure~\ref{fig:detection-lifecycle} illustrates the ML-based ransomware detection workflow. The process is divided into two key phases: the training and testing phase, and the protection phase. In the first phase, labeled datasets containing known benign software (benignware) and various ransomware families are used to train a predictive model capable of distinguishing between malicious and legitimate behavior. Once trained, the model transitions to the protection phase, where it evaluates unknown executables. Based on learned patterns, the model assigns a label —either ransomware or benignware —enabling real-time detection and prevention mechanisms. 

This problem is particularly relevant in large-scale, real-world deployments, such as those involving cybersecurity companies and vendors of software or hardware-based protection systems. Hardware vendors such as Microsoft~\cite{agrawal2019attention,microsoft2022improving,microsoft2017windows}, Google~\cite{huang2018tracking,oz2023rob,google2021we}, Cisco, NetApp, Fortinet, Palo Alto Networks, Versa Networks, Cloudflare, or Zscaler provide infrastructure and firewall appliances that monitor millions of customer environments, while software-based security companies like CrowdStrike, SentinelOne, Darktrace, ESET, Trellix, Sophos, or Kaspersky deliver endpoint protection solutions capable of detecting and mitigating ransomware across vast, distributed fleets of devices. These organizations typically have their detection agents or sensors deployed across millions of endpoints worldwide, yet they face stringent privacy, contractual, and regulatory restrictions that prohibit transferring local telemetry data or even security logs outside the customer’s environment.

As a result, traditional centralized ML approaches are infeasible. FL provides a privacy-preserving alternative, enabling these companies to collaboratively train robust, up-to-date detection models across distributed customer nodes without moving sensitive data from local environments. This allows global threat intelligence to emerge from local observations, bridging the gap between privacy and performance in large-scale ransomware detection.

The development and evaluation of malware detection models heavily rely on diverse and representative datasets~\cite{cic_malmem2022, dener2022malware, shafin2023obfuscated, cciplak2025fedetect, meidan2018n, anderson2018ember, freitas2022malnet, genccaydin2022benchmark}. These datasets provide the necessary data to train, validate, and benchmark ML models, ensuring their effectiveness in real-world scenarios.

\subsection{Related Work}

In the context of IoT environments, Rey et al.~\cite{rey2022federated} proposed a federated framework for malware detection that enables collaboration across resource-constrained devices while preserving user privacy. Their results demonstrate that FL can maintain high detection performance while avoiding centralized data collection. Similarly, Fang et al.~\cite{fang2023comprehensive} introduced a comprehensive Android malware detection system based on a federated architecture. Their approach integrates local models trained on device-specific data and aggregates them to build a robust global classifier, showing strong resistance to data heterogeneity. Emphasizing privacy, Galvez et al.~\cite{galvez2020less} developed a lightweight Android malware classifier that uses FL to respect user data confidentiality. Their model achieves competitive accuracy while reducing communication overhead, making it suitable for deployment on edge devices. These studies collectively highlight the viability of FL in malware detection tasks and support its extension to more targeted use cases such as ransomware detection, as explored in this work.

\subsection{Problem Description}
\label{sec:prob-description}
Let $N$ be the total number of rows (software samples) in the tabular dataset. For each software sample $i \in \{1, \dots, N\}$, we define a $p$-dimensional feature vector:
\[
\mathbf{x}_i = [x_{i,1}, x_{i,2}, \dots, x_{i,p}]^\top \in \mathbb{R}^p,
\]
The corresponding binary label for each row is:
\[
  y_i = \begin{cases}
    1, & \text{if row }i\text{ corresponds to a ransomware (malicious) sample},\\
    0, & \text{otherwise}.
  \end{cases}
\]
The complete dataset can be represented as:
\[
  \mathbf{X} = [\mathbf{x}_1, \mathbf{x}_2, \dots, \mathbf{x}_N]^\top \in \mathbb{R}^{N\times p}, \quad
  \mathbf{y} = [y_1, y_2, \dots, y_N]^\top \in \{0,1\}^N.
\]

This formulation yields a supervised binary classification task: given $\mathbf{x}_i$, predict $y_i$.

\section{ML privacy-preserving solution}

In this section, we provide a detailed explanation of the privacy-preserving ML solution, privacy and regulatory limitations, and an introduction to FL.

\subsection{The ML Approach}

Given an input space $\mathcal X\coloneqq \mathbb{R}^d$ and output set $\mathcal Y\subseteq \mathbb{R}^m$, the goal of supervised ML is roughly to approximate an unknown function, parameterized by $\theta$:
\begin{equation}\label{f}
	f_\theta:\mathcal{X}\longrightarrow \mathcal{Y},
\end{equation}
given a dataset $\mathcal D = \left\{\left(\mathbf{x}^{i}, \mathbf{y}^{i}\right)\right\}_{i=1}^N\subset \mathcal{X}\times \mathcal{Y}$, composed of $N$ known but possibly noisy examples, i.e.:
\begin{equation}
	\mathbf{y}^{i}\simeq f_\theta(\mathbf{x}^{i}).
\end{equation}

This approximation problem is typically formulated as the minimization of an Empirical Risk (ER) \cite{rumelhart1986learning,goodfellow2016deep} on some training data.

To that purpose (together with preprocessing) the dataset:
\begin{equation}
\mathcal{D}=\left\{\left(\mathbf{x}^{i},\mathbf{y}^{i}\right)\right\}_{i=1}^N 
\end{equation}
is firstly split into:
\begin{enumerate}
	\item[a)] training data
	\begin{equation}\label{centralzied_train_dataset}
	\mathcal{D}_{\mbox{\tiny{train}}}=\left\{\left(\mathbf{x}^{i},\mathbf{y}^{i}\right)\right\}_{i\in I_{\mbox{\tiny{train}}}};
	\end{equation}
	\item[b)] testing data
	\begin{equation}\label{centralzied_test_dataset}
	\mathcal{D}_{\mbox{\tiny{test}}}=\left\{\left(\mathbf{x}^{i},\mathbf{y}^{i}\right)\right\}_{i\in I_{\mbox{\tiny{test}}}};
	\end{equation}
	\item[c)] validation data
	\begin{equation}\label{}
	\mathcal{D}_{\mbox{\tiny{validation}}}=\left\{\left(\mathbf{x}^{i},\mathbf{y}^{i}\right)\right\}_{i\in I_{\mbox{\tiny{validation}}}},
	\end{equation}
\end{enumerate}
with
\begin{equation}
\left\{1,\dots,N\right\}=I_{\mbox{\tiny{train}}}\bigsqcup I_{\mbox{\tiny{test}}} \bigsqcup I_{\mbox{\tiny{validation}}}.
\end{equation}

At this point, for example, the empirical risk can be defined as:
\begin{equation}
	J: \mathbf{\Theta} \longrightarrow \mathbb{R}
\end{equation}
\begin{equation}\label{eq:functional}
J(\mathbf{\theta}) \coloneqq \frac{1}{\# I_{\mbox{\tiny{train}}}}\sum_{i\in I_{\mbox{\tiny{train}}}} \mathrm{loss}\Big(f_{\mathbf{\theta}}(\mathbf{x}^{i}), \mathbf{y}^{i}\Big) + \lambda \hspace{0.03 cm} \text{Reg}\left(\mathbf{\theta}\right),
\end{equation}
where:
\begin{itemize}
	\item The parameters space $\mathbf{\Theta}$ is a Hilbert space on $\mathbb{R}$;
	\item The continuous loss function:
	\begin{align*}
	\mathrm{loss}:\mathbb{R}^m\times \mathcal{Y}\longrightarrow \mathbb{R}^+
	\end{align*}
	penalizes the mismatch between the predictions $f_{\mathbf{\theta}}(\mathbf{x}^{i})$ and the labels $\mathbf{y}^{i}$;
	\item The \textit{regularization} term $\lambda$ penalizes the model overfitting on training data, the effect of this penalization being modulated by the weighting factor $\lambda>0$ and $\text{Reg}:\mathbf{\Theta}\longrightarrow \mathbb{R}^+$ being a function (e.g., $\text{Reg}\left(\mathbf{\theta}\right)=\left\|\mathbf{\theta}\right\|_{\mathbf{\Theta}}^2$ the squared Hilbertian norm);
	\item The model:
	\begin{equation}
		f_{\mathbf{\theta}}:\mathbb{R}^d\longrightarrow \mathbb{R}^m
	\end{equation}
	is a function, belonging to a class:
	\begin{align*}
	\mathscr{C}=\big\{f_{\mathbf{\theta}} \ | \ \mathbf{\theta} \in \mathbf{\Theta}\big\},
	\end{align*}
	$\mathbf{\theta}$ being the so-called trainable parameters; examples of $\mathscr{C}$ are Deep Neural Networks \cite{goodfellow2016deep}, Random Forest (RF) \cite{breiman2001random}, Gradient Boosted Decision Trees (GBDT) \cite{chen2016xgboost}, transformers \cite{vaswani2017attention}, Large Language Models \cite{naveed2025comprehensive} and Residual Neural Networks (ResNets) \cite{he2016deep}; $f_{\mathbf{\theta}}$ is designed to approximate \eqref{f}, for an appropriate choice of the parameters $\mathbf{\theta}$.
\end{itemize}

In the above context, the ML training is formulated as:
\begin{align}\label{opt_pb}
\mathbf{\theta}^\ast \in \underset{{\mathbf{\Theta}}}{\mbox{argmin}} J(\mathbf{\theta}).
\end{align}

\begin{remark}[Existence of solutions]
	Existence of a solution to Equation~\eqref{opt_pb} might be analyzed by the Direct Method in the Calculus of Variations \cite{dacorogna2007direct}.
	
	For instance, existence holds, assuming:
	\begin{itemize}
		\item The parameters space $\mathbf{\Theta}$ is of finite dimension;
		\item The regularization weighting parameter $\lambda >0$;
		\item The regularization function is the Hilbertian norm:
		\begin{equation}
			\text{Reg}\left(\mathbf{\theta}\right)=\left\|\mathbf{\theta}\right\|_{\mathbf{\Theta}}^2;
		\end{equation}
		\item For any $\mathbf{x}\in \mathbb{R}^d$, the function:
		\begin{equation}
			\mathbf{\theta}\mapsto f_{\mathbf{\theta}}(\mathbf{x})
		\end{equation}
		is continuous.
	\end{itemize}
\end{remark}

\begin{remark}[Convexity]
	$J$ might not be convex, even in case $\mathrm{loss}$ is convex. Indeed, convexity also depends on
	\begin{equation}
	\mathbf{\theta}\mapsto f_{\mathbf{\theta}}.
	\end{equation}
	
	In case $J$ is not strictly convex, even if a global minimizer exists, its uniqueness is not guaranteed.
\end{remark}

\subsection{Privacy and Regulatory Limitations}

Under the General Data Protection Regulation (GDPR)~\cite{gdpr_eurlex}, personal data is defined as any information related to an identified or identifiable natural person, known as the data subject. In the context of ransomware detection, system logs, user activity traces, file access records, and network traffic data often contain sensitive identifiers such as usernames, IP addresses, device IDs, or file paths that may link back to individuals or organizations. As such, these data fall within the scope of GDPR, and their unauthorized transmission or exposure can lead to serious legal and financial consequences.

Even when personal identifiers are pseudonymized or data is aggregated, uploading large volumes of raw behavioral logs or full activity traces imposes significant bandwidth and latency costs. This is particularly problematic in environments like enterprise endpoints, industrial systems, or edge devices with intermittent connectivity, making centralized ransomware detection impractical in many real-world deployments.

A more efficient and privacy-conscious alternative is to exchange model parameters or statistical summaries instead of raw data. Transmitting model updates or gradient information (typically in kilobytes or megabytes) is far more bandwidth-efficient and privacy-preserving than sharing raw logs. In this work, we employ HFL to collaboratively train ransomware detection models across distributed nodes without exposing local data. This approach ensures GDPR compliance while respecting hardware limitations and communication constraints. Moreover, by reducing the frequency and volume of data transmission, this setup aligns with \textit{Green ML} goals~\cite{mehta2023review}, helping minimize energy consumption and the environmental impact of distributed ML.

\subsection{Introduction to FL}
\label{sec:fedlearning-overview}

FL enables multiple \emph{nodes} to collaboratively train a global model without exchanging raw data~\cite{mcmahan2017communication}.  Instead, each node \(k\) maintains a local model \(f_{\theta_k}\) and periodically transmits model parameters or gradients to a central server (the \emph{aggregator}).  The server combines these updates into a global model and redistributes it to each node. This cycle repeats until convergence.

FL, while preserving data privacy and allowing edge devices to train with their data, introduces several challenges. The most notable one is related to handling non‐Independent and Identically Distributed (non‐IID) data across nodes, which can yield divergent local updates and degrade global accuracy. This problem is also referred to in the literature as \textit{data drift} or \textit{concept drift}~\cite{lu2024federated,zhu2021federated, zhao2018federated}.

We clarify that, in this paper, the focus is on handling non-IID scenarios. Rather than synthetically partitioning the data, we leverage the natural distribution present in the dataset, which contains information about distinct machines. Each machine is treated as an independent node, thereby creating a realistic federated setting where data distributions are non-IID across nodes (see Section~\ref{sec:nodes_creation}). 

\subsubsection{FL Paradigms}
\label{sec:fedlearning-paradigms}

Different data‐distribution scenarios give rise to distinct FL paradigms:
\begin{itemize}
  \item \textbf{HFL}:  In this paradigm, all nodes share the same feature space but possess different samples (rows).  \emph{Example}: multiple, disjoint banks collect transactions in the same way about their clients. 
  \item \textbf{Vertical FL (VFL)}: Nodes hold complementary feature subsets for the same samples (shared index set).  \emph{Example}: One bank records financial transactions, while another company records real estate acquisitions, such as purchases of buildings or companies.
\end{itemize}

In this work, we focus exclusively on HFL. After training is complete in an HFL setup, the resulting global model is typically shared with all participating parties. This allows each party to download the trained model and subsequently perform inference locally and independently, without requiring further interaction or data exchange.

\subsubsection{HFL}
\label{sec:fedlearning-hfl}

Under HFL, each node \(k\) has a local dataset:
\[
\mathcal{D}_k
= \bigl\{(\mathbf{x}_i^k,\,\mathbf{y}_i^k)\bigr\}_{i=1}^{N_k},
\]
where all \(\mathbf{x}_i^k\in\mathbb{R}^p\) share the same feature dimension \(p\), but the number of samples \(N_k\) can differ.  Training proceeds as follows:
\begin{enumerate}
  \item \textbf{Local update:} Each node \(k\) minimizes its empirical risk function until local convergence.
\[
J_k(\theta)
= \frac{1}{N_k}
  \sum_{i=1}^{N_k}
    \mathcal{L}\bigl(f_\theta(\mathbf{x}_i^k),\,\mathbf{y}_i^k\bigr)
\]
  \item \textbf{Aggregation:} Nodes send their updated parameters \(\theta_k\) to the server.
  \item \textbf{Global update:} The server aggregates the sets of local updated parameters \(\{\theta_k\}\) into a new global federated set of parameters \(\theta\).
  \item \textbf{Broadcast:} The server distributes \(\theta\) back to all nodes and the process starts again.
\end{enumerate}

\section{Centralized Dataset and Preprocessing}

In this section, we describe the dataset, outline the preprocessing steps, and present the centralized architecture.

\subsection{Description of the Dataset}
\label{sec: description}

The choice of appropriate datasets is crucial for training effective malware detection models. We selected the ransomware storage access
patterns (RanSAP) dataset~\cite{hirano2022ransap, hirano2019machine} due to its comprehensive coverage of both benign and malicious samples, including a wide variety of malware types, making it suitable for evaluating real-world detection capabilities. Additionally, its per-device structure aligns well with HFL, as multiple endpoints associated with cybersecurity companies' customers (in this case, represented by their servers) can contribute similar feature representations without sharing raw binaries or system logs, thus preserving data privacy. 

\begin{figure}[h]
    \centering
    \includegraphics[width=0.95\linewidth]{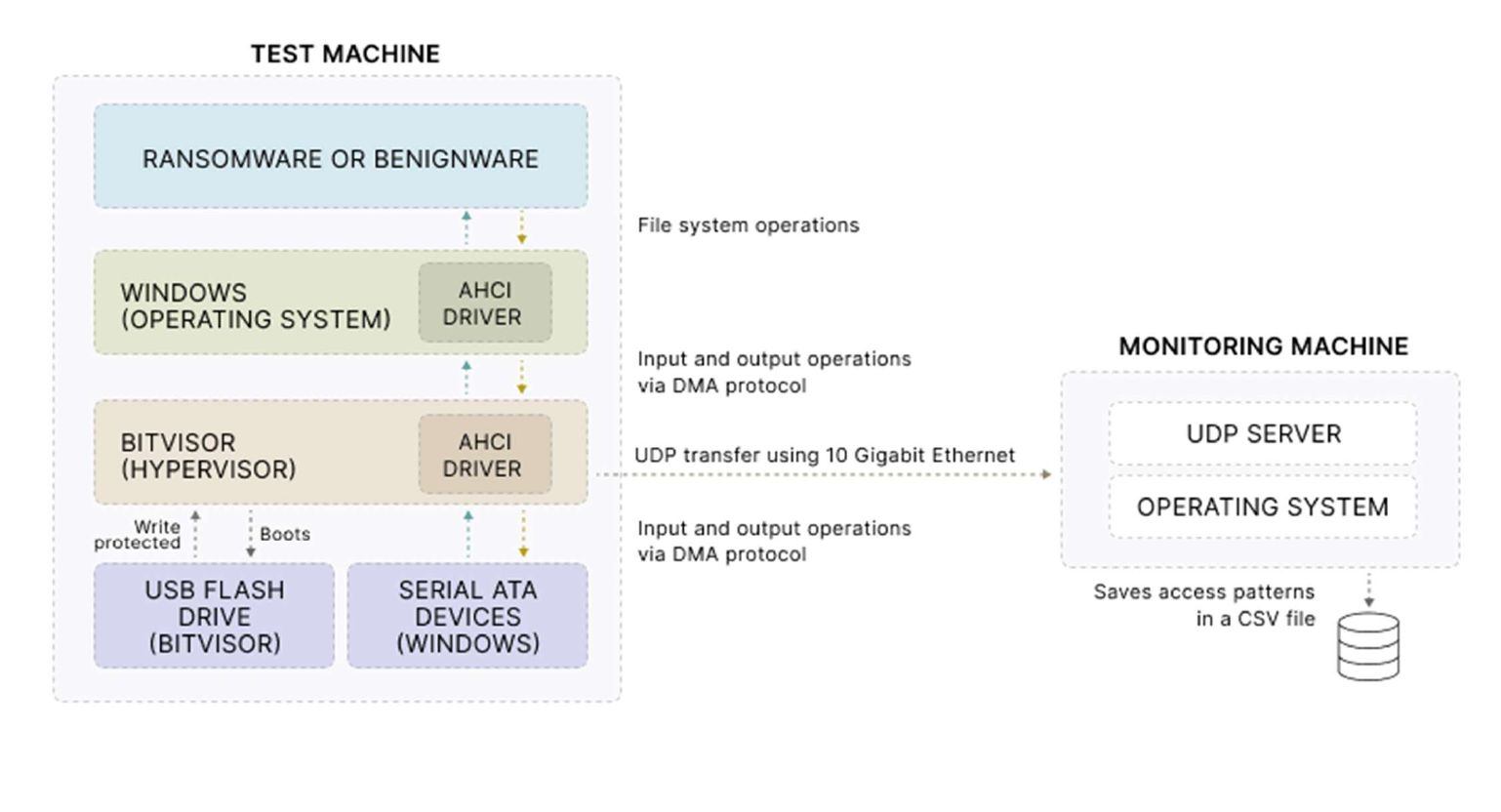}
    \caption{Overview of the RanSAP data collection environment. A write-protected USB containing BitVisor boots the test machine. The hypervisor intercepts AT Bus Attachment (ATA) input/output operations between the Windows OS and the storage device using the Advanced Host Controller Interface (AHCI) protocol. These access patterns are transmitted via a 10 Gbps Ethernet connection using User Datagram Protocol (UDP) to a monitoring machine, which records them as CSV files. Image modified from Hirano et al.~\cite{hirano2022ransap}}
    \label{fig:ransap-architecture}
\end{figure}

The dataset aims to \textit{detect the presence or absence of ransomware} on four servers, all running Windows 7, but differing in memory type and capacity: two use hard disk drives (HDDs) (120 GB and 250 GB) and two use solid-state drives (SSDs) (120 GB and 250 GB). Each system contains a variety of installed software, both benign and malicious, with some ransomware samples associated with decoy files, intentionally crafted files designed to detect ransomware activity. These decoys are executed in two modes: \texttt{-largefiles} (large files such as \texttt{.ppt, .txt, .xls, .doc,} and \texttt{.ps}) and \texttt{-w10dirs} (mimicking Windows 10 directory structures, with file types like \texttt{.pdf, .html, .txt, .doc,} and \texttt{.ppt}). To further clarify the data collection mechanism, Figure~\ref{fig:ransap-architecture} illustrates the RanSAP experimental setup. Ransomware, which encrypts user data and demands payment for decryption, is the main focus. The software used is grouped by type rather than by the presence of decoys. Benign software (labeled zeros) includes \texttt{AESCrypt, Zip, SDelete, Excel,} and \texttt{Firefox}, while malicious software (labeled ones) includes \texttt{TeslaCrypt, Cerber, WannaCry, GandCrab, Ryuk, Sodinokibi,} and \texttt{Darkside}, each with its specific behavior and infection mechanisms. For each software, 10–11 date-stamped folders contain two CSV files (ata\_read.csv and ata\_write.csv) representing memory read and write operations. These were captured using BitVisor, a hypervisor booted from a write-protected USB to avoid compromise. The OS is launched from a test HDD or SSD, where ransomware or benign samples are executed. The AHCI interface intercepts low-level memory I/O via DMA, and access patterns are transmitted via UDP to a monitoring machine. The resulting CSV files contain detailed logs of memory usage, reflecting software behavior in terms of read and write access patterns.

\subsection{Preprocessing of the Dataset}

The raw dataset consists of two CSV files containing records each time a ransomware sample or benign software sample is executed. The first file, named \texttt{ata\_read.csv}, represents the data matrix $\mathbf{R} = \{(t_i^{(s)}, t_i^{(\mu)}, l_i^{(r)}, b_i^{(r)})\}_{i=1}^{n_r}$. The second CSV, \texttt{ata\_write.csv}, records similar information $\mathbf{W} = \{(t_j^{(s)}, t_j^{(\mu)}, l_j^{(w)}, b_j^{(w)}, e_j^{(w)})\}_{j=1}^{n_w}$. Here we denote: 
\begin{itemize}
  \item $t_i^{(s)}, t_j^{(s)} \in \mathbb{N}$ are timestamps in seconds,
  \item $t_i^{(\mu)}, t_j^{(\mu)} \in \mathbb{N}$ are timestamps in microseconds,
  \item $l_i^{(r)}, l_j^{(w)} \in \mathbb{N}$ are Logical Block Addresses (LBAs) for read and write events, which specify the location of blocks on an ATA device such as an HDD or SSD
  \item $b_i^{(r)}, b_j^{(w)} \in \mathbb{N}$ are the sizes of read and written blocks in bytes,
  \item $e_j^{(w)} = -\sum_{i=1}^n p_i \log_2(p_i)/\log_2(n) \in [0, 1]$ is the normalized Shannon entropy of written data, where $p_i$ is a probability of a byte $i$, which is an $i$-th byte in a sector $s$, and $n$ is the size of a sector in bytes, in our case, 512 bytes. 
\end{itemize}

However, these vectors alone do not reflect changes in access patterns over time. If all vectors were used separately, the resulting vector space would be too similar, leading to poor performance metrics. To address this, five features forming the data vector $\mathbf{x}_k$ are derived from the original two CSV files using a moving average within a fixed time window $T = 30s$: 
\begin{align*}
  \mathrm{AvgEntropyWrite}_k &= \frac{1}{|\mathcal{W}_k|} \sum_{j \in \mathcal{W}_k} e_j^{(w)}, \\
  \mathrm{VarLBAWrite}_k &= \mathrm{Var}(\{l_j^{(w)} \mid j \in \mathcal{W}_k\}), \\
  \mathrm{AvgWriteThroughput}_k &= \frac{1}{T} \sum_{j \in \mathcal{W}_k} b_j^{(w)}, \\
  \mathrm{VarLBARead}_k &= \mathrm{Var}(\{l_i^{(r)} \mid i \in \mathcal{R}_k\}), \\
  \mathrm{AvgReadThroughput}_k &= \frac{1}{T} \sum_{i \in \mathcal{R}_k} b_i^{(r)},
\end{align*}
with $\mathcal{W}_k$ and $\mathcal{R}_k$ denoting indices of write and read events falling within the $k$-th time window. Each sample $k$ is labeled with $y_k \in \{0, 1\}$, where $y_k = 1$ indicates ransomware and $y_k = 0$ indicates benign behavior. The dataset for a given node $m$ is then:
\[
\mathcal{D}_m = \{(\mathbf{x}_k^{(m)}, y_k^{(m)})\}_{k=1}^{N_m},
\]
where $N_m$ is the number of time windows for node $m$. The centralized dataset is created by merging the data across different computers (servers) for centralized analysis, while datasets per individual node are formed for naturally \textit{heterogeneous} HFL (explained in more detail in Section~\ref{sec:fed_arch}). The data was then treated with general preprocessing procedures on the basis of the exploratory data analysis (data normalization and class balancing with the combination of under- and oversampling). The overall goal is to collaboratively learn a global classifier $f_\theta(\mathbf{x}): \mathbb{R}^5 \rightarrow \{0,1\}$ parameterized by some set $\theta$.

\subsection{Centralized Architecture}
\label{sec: cen-architecture}

\begin{figure}[h!]
  \centering
  \includegraphics[scale=0.7]{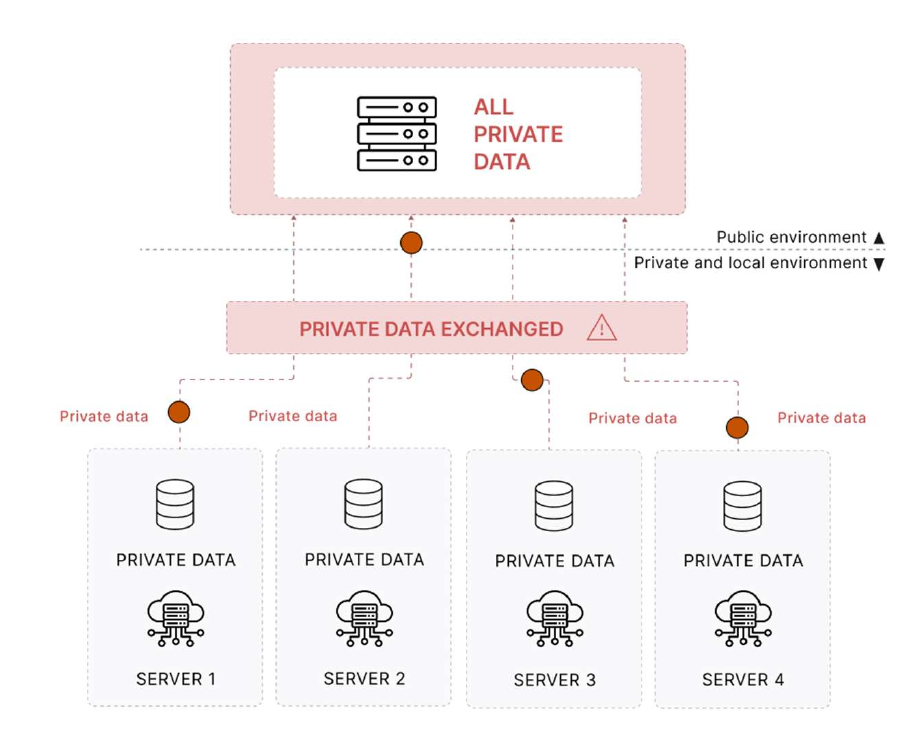}
  \caption{\label{fig:cent_arch} Centralized architecture.}
\end{figure}

Let us first examine two limiting cases for our experiments. The lower benchmark limit for the trained model can be obtained by each single node (in our case \textendash{ }server) training a model on its own, on its local dataset, assuming no data can be shared with other nodes. In what follows, we refer to this as the \textit{local} training. Here we expect to attain the baseline performing models, due to a lack of training data and its natural heterogeneity across the servers.

Another case, the best possible limit, is the so-called \textit{centralized} training (see Figure~\ref{fig:cent_arch}), a training scenario in which all the nodes have been allowed to join together their datasets and train a unique model. The reasons preventing this are also discussed in Section~\ref{sec:fed_arch}.
For the ransomware detection task using the RanSAP dataset, we selected the RF as the primary classification model. The same model was used in the original paper~\cite{hirano2022ransap}, so that we can directly compare the results.

RF~\cite{breiman2001random} is an ensemble learning method that constructs a multitude of decision trees during training and outputs their averaged predictions. The algorithm introduces two primary sources of randomness: bootstrapping of the training data (bagging) and random feature selection at each split. This dual-randomization reduces variance and guards against overfitting, particularly in high-dimensional feature spaces.

RF is well-regarded for its robustness to noise, ability to handle missing data, and interpretability via feature importance measures. These properties are essential in cybersecurity applications, where datasets may be incomplete or noisy, and model explainability is often critical for incident response. Besides, the \sherpa FL platform features an implementation for the Federated RF model that is extremely communication-efficient, requiring only two communication rounds between the node and the Platform, yet preserving all its advantages and flexibility. The ability to model nonlinear interactions and hierarchical feature relationships makes them particularly effective for malware and ransomware detection tasks, where behavioral signatures may be subtle and context-dependent.
\section{Proposed privacy-preserving Solution through FL}
\label{sec:fed_arch}
In traditional ML, all training data must be collected and centralized in a single location before model training can begin. This requires data from different silos, such as various organizations or departments, to be transferred to a central server. Such aggregation (upon which we have already mentioned in Section~\ref{sec: cen-architecture}, and see also Figure~\ref{fig:cent_arch}) introduces several significant limitations. First, when the data involved is sensitive, as is often the case in cybersecurity applications like malware detection, transferring it to a central repository can violate data protection regulations such as \cite{gdpr2016}, \cite{lopd2018}, \cite{hipaa1996}, or \cite{CCPA2018}. Furthermore, once data is shared, data owners lose control over it, and it becomes vulnerable during transmission and storage. The centralized approach, while being an ideal-world benchmark training scenario and thus, hypothetically, providing the best-performing model, also creates a single point of failure, increasing the risk of privacy breaches and compromising the security of the entire dataset.

FL holds significant promise for improving ransomware detection by enabling collective learning across isolated datasets. In view of the problem, consider a scenario involving several financial institutions or healthcare providers. Each organization independently collects behavioral telemetry indicative of ransomware, such as provided in the RanSAP dataset. However, due to confidentiality constraints, raw data (in our concrete simulated case{ }\textendash{ }the separate servers's data) cannot be shared. Through HFL~\cite{konevcny2016federated, konevcny2016federated2, wang2021field} these organizations can collaboratively train a detection model that benefits from a diverse set of ransomware behaviors across environments, thereby improving generalizability and detection performance.

Smart manufacturing provides another relevant example. Factories employing IoT-enabled devices such as computer numerical control (CNC) machines and industrial robots face growing threats from ransomware targeting operational technology (OT). These environments generate high-volume telemetry that is critical for early anomaly detection but often includes sensitive operational details. By leveraging FL, factories can retain proprietary data while contributing to a shared ransomware detection model, capturing early indicators of compromise, such as unauthorized encryption or anomalous access to programmable logic controllers (PLCs), without disclosing the contents or structure of their control systems.

\subsection{FL Architecture}

\begin{figure}[htpb]
  \centering
  \includegraphics[width=0.6\textwidth]{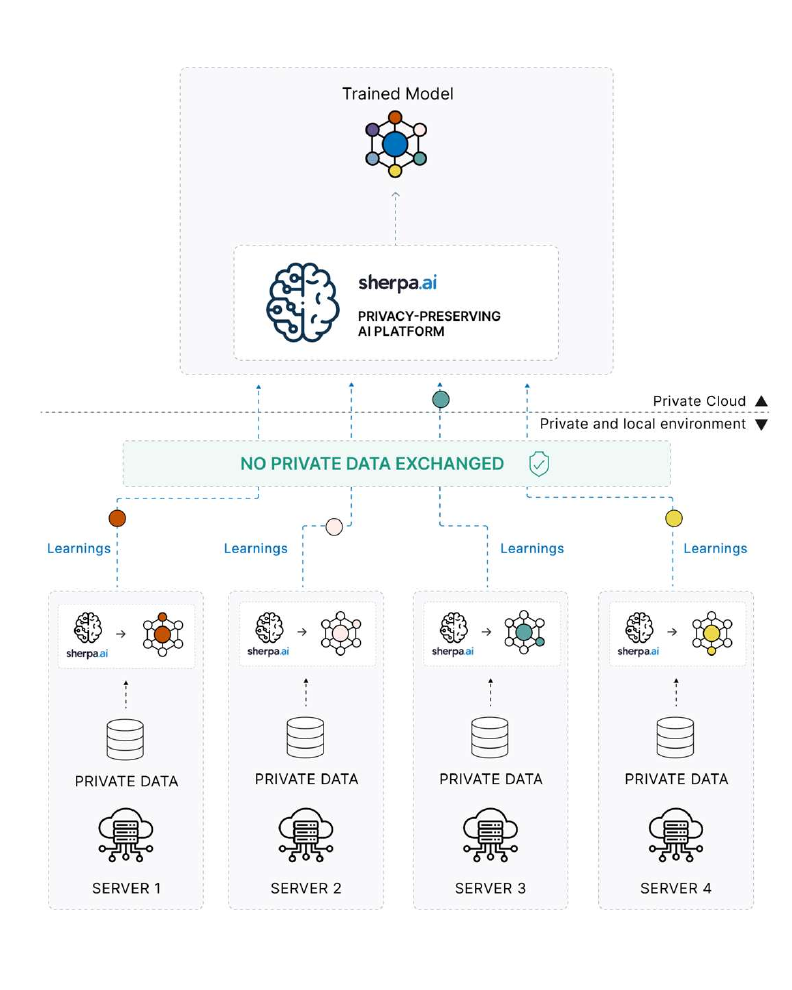}
    \caption{\label{fig:centfed} Federated architecture implemented on the \sherpa FL platform.}
\end{figure}

FL offers a decentralized alternative in which data remains within its original silo. Using HFL, as implemented on the \sherpa FL platform (see Figure~\ref{fig:centfed}), a global model is trained by sending initial model parameters to each local node, where training occurs on the local data. The \textit{locally computed updates} (i.e., gradients or model weights) are then sent back and aggregated centrally to update the global model. This method ensures that raw data never leaves its origin, enabling privacy-preserving collaborative training. It aligns with regulatory requirements and minimizes the risk of data breaches by reducing the attack surface and eliminating the need for direct data sharing. This paradigm allows entities to collaboratively train robust malware detection models while retaining full control over their data and maintaining compliance with data protection standards.

The results of the three experiments { }\textendash{ } the centralized, federated, and local (by nodes) { }\textendash{ } are discussed below.

\subsection{Creation of Nodes} \label{sec:nodes_creation}

The RanSAP dataset comprises behavioral data from four servers, resulting in a federated setting with four servers that simulate distinct customer environments. This setup is consistent with deployments where agents operate across millions of endpoints, but security event logs cannot be exported outside the customer’s infrastructure. In our approach, each server's data is treated as the dataset of an individual node, preserving the inherent distribution differences across systems. For each node, we held out 25\% of samples via stratified random sampling to form the test set (implemented using scikit-learn's \texttt{train\_test\_split(stratify=y, test\_size=0.25)}). No separate validation set was used.

\begin{table}[htpb]
  \scriptsize
  \setlength{\tabcolsep}{4pt}             
  \renewcommand{\arraystretch}{1.2}
  \centering
  \begin{tabularx}{0.5\textwidth}{L|Y|Y}      
    \toprule
    \textbf{Server} & $N_m$ \textbf{(train dataset length)} & \textbf{Test dataset length}\\
    \midrule
    win7-120gb-hdd & 11940 & \multirow{4}{*}{15923} \\
    win7-120gb-ssd & 11895 \\
    win7-250gb-hdd & 11986 \\
    win7-250gb-ssd & 11940 \\
    \midrule
    \textbf{Total} & \textbf{47761} & \textbf{15923} \\
    \bottomrule
  \end{tabularx}
  \vspace{5mm}
  \caption{Datasets lengths for different servers (nodes).}
  \label{tab:dataset_lengths}
\end{table}

Table \ref{tab:dataset_lengths} summarizes the number of examples used for training and testing across the four servers (nodes) in the federated setting. Each server maintains a comparable amount of training data, while the centralized test set, containing 15,923 examples, is used to evaluate the experiments defined in Section~\ref{sec:experiments}. The dataset comprises a total of 63,384 samples, distributed among the four servers.

\section{Experiment} \label{sec:experiments}
In this section, we detail the experimental setup, including the evaluation metrics, training and testing configurations, and the main results.

We performed a set of experiments using the scikit-learn~\cite{scikit-learn} RF model implementation with default parameters (as suggested in~\cite{hirano2022ransap}), under the following training scenarios:
\begin{enumerate}
    \item \textbf{Centralized}: All client datasets were merged into a single dataset; the RF model was trained centrally.
    \item \textbf{Federated}: Each local node trained its RF model on its own dataset while participating in the federated process.
    \item \textbf{Single-node}: Each node trained its own RF model in isolation on its local dataset.
\end{enumerate}

In all cases, a unified test dataset was constructed by aggregating test data from each of the participating nodes, ensuring a representative evaluation of generalization across the entire data distribution of the servers.

\subsection{Evaluation Metrics}
\label{sec:metrics}
To assess the performance of the proposed experiments, we employ standard classification metrics: Accuracy, Precision, Recall, and F1-score. These metrics are computed based on the confusion matrix, which consists of True Positives (TP), False Positives (FP), True Negatives (TN), and False Negatives (FN).

\begin{itemize}
    \item \textbf{Accuracy} measures the proportion of correctly classified samples among all samples. Although widely used, it can be misleading in imbalanced datasets.
    \begin{equation}
        \text{Accuracy} = \frac{TP + TN}{TP + TN + FP + FN}
    \end{equation}

    \item \textbf{Precision} indicates the proportion of predicted positive samples that are actually positive. In malware detection, this reflects the rate of correctly identified malware among all predicted malware.
    \begin{equation}
        \text{Precision} = \frac{TP}{TP + FP}
    \end{equation}

    \item \textbf{Recall} also known as \textbf{sensitivity} or \textbf{true positive rate}, quantifies the proportion of actual positives that are correctly identified. This is crucial in malware detection, where failing to identify malware (false negatives) can be costly.
    \begin{equation}
        \text{Recall} = \frac{TP}{TP + FN}
    \end{equation}

    \item \textbf{F1-score} is the harmonic mean of Precision and Recall. It provides a balanced measure that accounts for both false positives and false negatives.
    \begin{equation}
        \text{F1-score} = 2 \cdot \frac{\text{Precision} \cdot \text{Recall}}{\text{Precision} + \text{Recall}}
    \end{equation}
\end{itemize}

Precision and Recall are particularly critical in this domain due to the asymmetric costs of false positives and false negatives.

\subsection{Experimentation Testbed}

All experiments were conducted using a machine equipped with 1 TB of disk space, an Intel Core i7-7700 4-core CPU at 3.60 GHz, 64 GB of RAM, the Ubuntu 24.04 operating system, and Python 3.11.

\subsection{Results}

\begin{figure*}[htbp]
  \centering
  \begin{subfigure}[t]{0.24\textwidth}
    \centering
    \includegraphics[width=\linewidth]{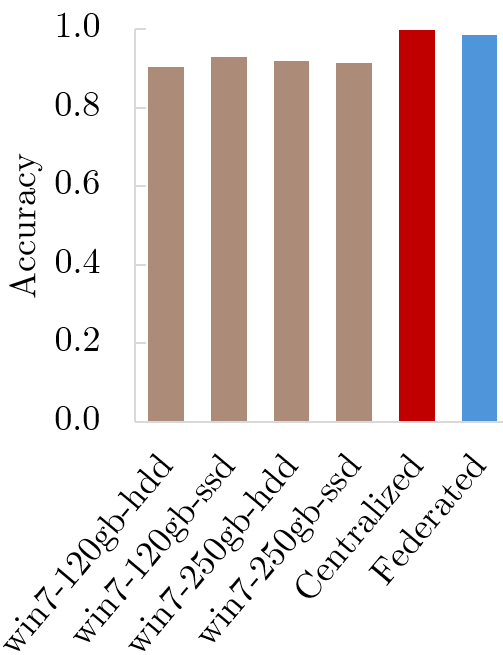}
    \caption{Accuracy}
    \label{fig:res-acc}
  \end{subfigure}\hfill
  \begin{subfigure}[t]{0.24\textwidth}
    \centering
    \includegraphics[width=\linewidth]{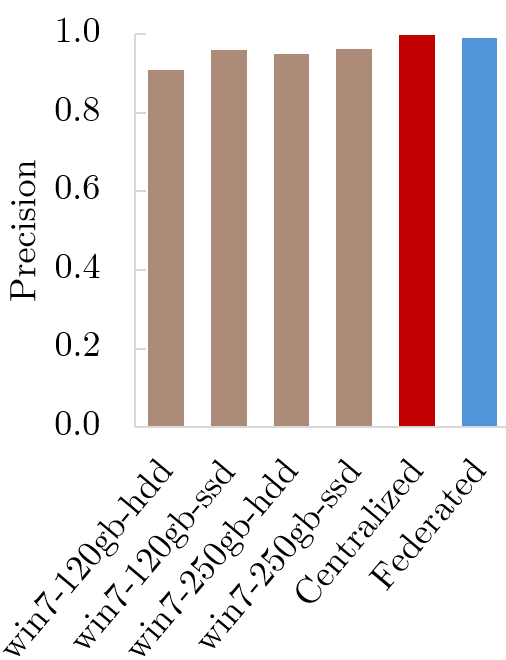}
    \caption{Precision}
    \label{fig:res-prec}
  \end{subfigure}\hfill
  \begin{subfigure}[t]{0.24\textwidth}
    \centering
    \includegraphics[width=\linewidth]{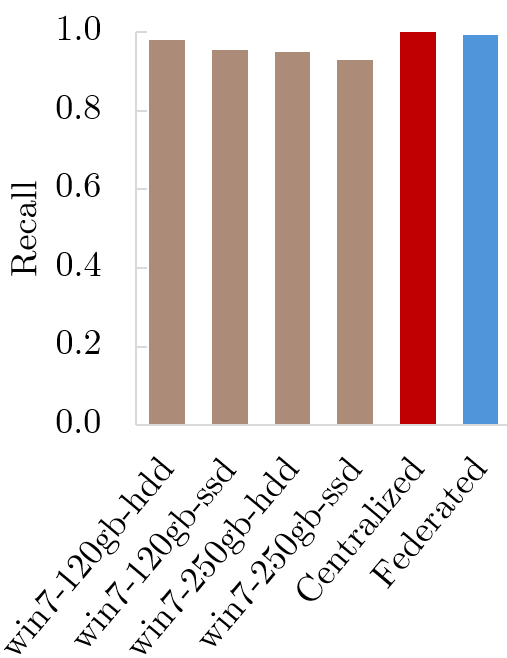}
    \caption{Recall}
    \label{fig:res-rec}
  \end{subfigure}\hfill
  \begin{subfigure}[t]{0.24\textwidth}
    \centering
    \includegraphics[width=\linewidth]{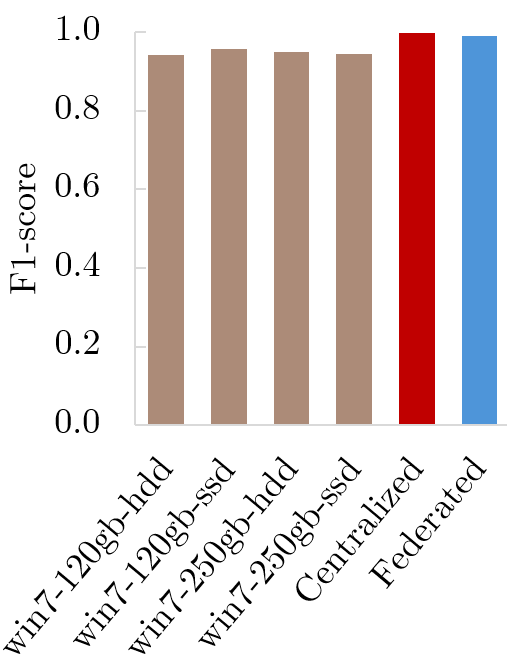}
    \caption{F1-score}
    \label{fig:res-f1}
  \end{subfigure}
  \caption{Metric values of the models' performances for the single-server models compared to the centralized and federated models.}
  \label{fig:res}
\end{figure*}

Table~\ref{tab:performance_comparison} reports the performance metrics (Section~\ref{sec:metrics}) for the RF models trained on individual servers (win7-120gb-hdd, win7-120gb-ssd, win7-250gb-hdd, win7-250gb-ssd), as well as for the FL and centralized models. Figure~\ref{fig:res} provides the corresponding visualization.



\begin{table}[htpb]
  \scriptsize
  \setlength{\tabcolsep}{4pt}             
  \renewcommand{\arraystretch}{1.2}
  \centering
  \begin{tabularx}{\textwidth}{L|Y|Y|Y|Y|Y|Y}      
    \toprule
        & \textbf{win7-120gb-hdd} & \textbf{win7-120gb-ssd} & \textbf{win7-250gb-hdd} & \textbf{win7-250gb-ssd} & \textbf{Centralized} & \textbf{Federated} \\ 
    \midrule
    Accuracy   & 0.905 & 0.930 & 0.919 & 0.913 & 0.999 & 0.986 \\ 
    \midrule
    Precision  & 0.908 & 0.960 & 0.950 & 0.962 & 0.999 & 0.990 \\ 
    \midrule
    Recall     & 0.981 & 0.954 & 0.950 & 0.929 & 1.000 & 0.992 \\ 
    \midrule
    F1-Score   & 0.943 & 0.957 & 0.950 & 0.945 & 0.999 & 0.991 \\ 
    \bottomrule
  \end{tabularx}
  \vspace{5mm}
  \caption{Mean metrics across single-servers (training on the node's local dataset), centralized model, and federated model, all evaluated on the same centralized test dataset.}
  \label{tab:performance_comparison}
\end{table}


As shown in Table~\ref{tab:performance_comparison} and Figure~\ref{fig:res}, models trained solely on local servers exhibited low to moderate performance, reflecting heterogeneous data distributions across nodes. The FL model outperformed every single-node model on all metrics, achieving a 9\% relative accuracy gain over the lowest-performing local model, demonstrating the benefit of collaborative training without sharing raw data. As expected, FL trailed the centralized model slightly, and the centralized model achieved the highest scores on all metrics.

The reason for the gap in performance (federated \textit{vs} centralized) is the architecture of the Federated RF model implemented in the \sherpa FL platform. Being optimized for speed, it reduces the communications overhead to only two rounds and the overall wall time for this specific training to less than 2 minutes. We also note that the F1-score obtained in our centralized and federated experiments aligns well with the results reported by Hirano et al.~\cite{hirano2022ransap} (see Figures 5 and 6 of the cited paper).

\section{Conclusions} \label{sec:Conclusions}

 
 

The experiments presented in this work confirm the effectiveness of the \sherpa FL platform for malware and ransomware detection across distributed data servers. The federated model achieves performance comparable to the centralized approach and clearly surpasses single-server baselines, all while maintaining strict data privacy. This framework is especially relevant for cybersecurity companies, including both software and hardware vendors, that operate at the scale of millions of endpoints. where contractual, regulatory, and privacy restrictions prevent the transfer of on-device logs beyond client environments.

Our study demonstrates that collaborative learning can be achieved without compromising sensitive data: organizations can jointly model and detect emerging ransomware behaviors in a privacy-preserving and regulatory-compliant threat intelligence.

Furthermore, the FL platform integrates seamlessly into existing ML workflows, and after collaborative training is completed, the resulting global model can be distributed and deployed locally, just as in a conventional system.
In summary, the proposed approach enables cybersecurity companies to reach high-accuracy, privacy-preserving advanced threat detection under real-world regulatory constraints, ensuring that critical customer information remains fully protected within its local environment.









\section*{Contributions and Acknowledgments} \label{app:A}


Daniel M. Jimenez-Gutierrez

Enrique Zuazua

Joaquin Del Rio

Oleksii Sliusarenko

Xabi Uribe-Etxebarria

\vspace{5mm}
The authors are presented in alphabetical order by first name. 

\printbibliography 

@article{rumelhart1986learning,
  title={Learning representations by back-propagating errors},
  author={Rumelhart, David E and Hinton, Geoffrey E and Williams, Ronald J},
  journal={Nature},
  volume={323},
  number={6088},
  pages={533--536},
  year={1986},
  publisher={Nature Publishing Group UK London}
}

@book{goodfellow2016deep,
  title={Deep learning},
  author={Goodfellow, Ian and Bengio, Yoshua and Courville, Aaron and Bengio, Yoshua},
  volume={1},
  number={2},
  year={2016},
  publisher={MIT press Cambridge}
}

@article{naveed2025comprehensive,
  title={A comprehensive overview of large language models},
  author={Naveed, Humza and Khan, Asad Ullah and Qiu, Shi and Saqib, Muhammad and Anwar, Saeed and Usman, Muhammad and Akhtar, Naveed and Barnes, Nick and Mian, Ajmal},
  journal={ACM Transactions on Intelligent Systems and Technology},
  volume={16},
  number={5},
  pages={1--72},
  year={2025},
  publisher={ACM New York, NY}
}

@inproceedings{he2016deep,
  title={Deep residual learning for image recognition},
  author={He, Kaiming and Zhang, Xiangyu and Ren, Shaoqing and Sun, Jian},
  booktitle={Proceedings of the IEEE conference on computer vision and pattern recognition},
  pages={770--778},
  year={2016}
}

@article{vaswani2017attention,
  title={Attention is all you need},
  author={Vaswani, Ashish and Shazeer, Noam and Parmar, Niki and Uszkoreit, Jakob and Jones, Llion and Gomez, Aidan N and Kaiser, {\L}ukasz and Polosukhin, Illia},
  journal={Advances in neural information processing systems},
  volume={30},
  year={2017}
}

@book{dacorogna2007direct,
  title={Direct methods in the calculus of variations},
  author={Dacorogna, Bernard},
  volume={78},
  year={2007},
  publisher={Springer Science \& Business Media}
}

@misc{gdpr2016,
  title = {Regulation (EU) 2016/679 of the European Parliament and of the Council of 27 April 2016 on the protection of natural persons with regard to the processing of personal data and on the free movement of such data, and repealing Directive 95/46/EC (General Data Protection Regulation)},
  journaltitle = {Official Journal of the European Union},
  date = {2016-04-27},
  volume = {L119},
  pages = {1--88},
  pagination = {page},
  number = {L 119/1},
  url = {https://eur-lex.europa.eu/legal-content/EN/TXT/PDF/?uri=CELEX:32016R0679}
}

@misc{hipaa1996,
  title = {Health Insurance Portability and Accountability Act of 1996},
  shorttitle = {HIPAA},
  number = {Pub. L. No. 104-191},
  pages = {110 Stat. 1936},
  date = {1996-08-21},
  pagination = {section}
}

@misc{CCPA2018,
  title = {California Consumer Privacy Act},
  journaltitle = {California Civil Code},
  date = {2018},
  number = {1798.100-1798.199},
  pagination = {section},
}

@misc{lopd2018,
  title = {Ley Orgánica 3/2018, de 5 de diciembre, de Protección de Datos Personales y garantía de los derechos digitales},
  journaltitle = {Boletín Oficial del Estado},
  date = {2018-12-06},
  number = {294},
  pages = {119788-119857},
  url = {https://www.boe.es/eli/es/lo/2018/12/05/3},
}

@article{konevcny2016federated,
	title={Federated optimization: Distributed machine learning for on-device intelligence},
	author={Kone{\v{c}}n{\`y}, Jakub and McMahan, H Brendan and Ramage, Daniel and Richt{\'a}rik, Peter},
	journal={arXiv preprint:1610.02527},
	year={2016}
}

@article{konevcny2016federated2,
	title={Federated learning: Strategies for improving communication efficiency},
	author={Kone{\v{c}}n{\`y}, Jakub and McMahan, H Brendan and Yu, Felix X and Richt{\'a}rik, Peter and Suresh, Ananda Theertha and Bacon, Dave},
	journal={arXiv preprint:1610.05492},
	year={2016}
}

@article{wang2021field,
	title={A field guide to federated optimization},
	author={Wang, Jianyu and Charles, Zachary and Xu, Zheng and Joshi, Gauri and McMahan, H Brendan and Al-Shedivat, Maruan and Andrew, Galen and Avestimehr, Salman and Daly, Katharine and Data, Deepesh and others},
	journal={arXiv preprint arXiv:2107.06917},
	year={2021}
}

@article{hirano2022ransap,
  title={RanSAP: An open dataset of ransomware storage access patterns for training machine learning models},
  author={Hirano, Manabu and Hodota, Ryo and Kobayashi, Ryotaro},
  journal={Forensic Science International: Digital Investigation},
  volume={40},
  pages={301314},
  year={2022},
  publisher={Elsevier},
  doi = {https://doi.org/10.1016/j.fsidi.2021.301314},
}

@inproceedings{hirano2019machine,
  title={Machine learning based ransomware detection using storage access patterns obtained from live-forensic hypervisor},
  author={Hirano, Manabu and Kobayashi, Ryotaro},
  booktitle={2019 sixth international conference on internet of things: Systems, Management and security (IOTSMS)},
  pages={1--6},
  year={2019},
  organization={IEEE}
}

@misc{techtarget_definition,
  author = {{TechTarget Editorial}},
  title = {Malware (malicious software)},
  year = {2022},
  note = {\url{https://www.techtarget.com/searchsecurity/definition/malware}}
}

@article{obfuscation_limitation,
  author = {Saxe, Joshua and Berlin, Konstantin},
  title = {Deep neural network based malware detection using two dimensional binary program features},
  journal = {10th International Conference on Malicious and Unwanted Software (MALWARE)},
  year = {2015},
  pages = {11--20}
}

@article{aslan2020comprehensive,
  title={A comprehensive review on malware detection approaches},
  author={Aslan, {\"O}mer Aslan and Samet, Refik},
  journal={IEEE access},
  volume={8},
  pages={6249--6271},
  year={2020},
  publisher={IEEE}
}

@article{meidan2018n,
  title={N-baiot—network-based detection of iot botnet attacks using deep autoencoders},
  author={Meidan, Yair and Bohadana, Michael and Mathov, Yael and Mirsky, Yisroel and Shabtai, Asaf and Breitenbacher, Dominik and Elovici, Yuval},
  journal={IEEE Pervasive Computing},
  volume={17},
  number={3},
  pages={12--22},
  year={2018},
  publisher={IEEE}
}

@inproceedings{mcmahan2017communication,
  title={Communication-Efficient Learning of Deep Networks from Decentralized Data},
  author={McMahan, H Brendan and Moore, Eider and Ramage, Daniel and Hampson, Seth and Arcas, Blaise Aguera y},
  booktitle={Proceedings of the 20th International Conference on Artificial Intelligence and Statistics (AISTATS)},
  pages={1273--1282},
  year={2017},
  organization={PMLR},
  url={https://proceedings.mlr.press/v54/mcmahan17a.html}
}

@misc{cic_malmem2022,
  author = {{Canadian Institute for Cybersecurity}},
  title = {CIC-MalMem-2022 Dataset},
  year = {2022},
  note = {\url{https://www.unb.ca/cic/datasets/malmem-2022.html}}
}

@article{dener2022malware,
  author = {Dener, M. and others},
  title = {Malware Detection Using Memory Analysis Data in Big Data Environment},
  journal = {Applied Sciences},
  volume = {12},
  number = {17},
  pages = {8604},
  year = {2022},
  publisher = {MDPI},
  doi = {10.3390/app12178604}
}

@article{shafin2023obfuscated,
  title={Obfuscated memory malware detection in resource-constrained IoT devices for smart city applications},
  author={Shafin, Sakib Shahriar and Karmakar, Gour and Mareels, Iven},
  journal={Sensors},
  volume={23},
  number={11},
  pages={5348},
  year={2023},
  publisher={MDPI}
}

@article{anderson2018ember,
  author = {Anderson, Hyrum S. and Roth, Phil},
  title = {EMBER: An Open Dataset for Training Static PE Malware Machine Learning Models},
  journal = {arXiv preprint arXiv:1804.04637},
  year = {2018}
}

@inproceedings{freitas2022malnet,
  title={MalNet: A large-scale image database of malicious software},
  author={Freitas, Scott and Duggal, Rahul and Chau, Duen Horng},
  booktitle={Proceedings of the 31st ACM International Conference on Information \& Knowledge Management},
  pages={3948--3952},
  year={2022}
}

@inproceedings{genccaydin2022benchmark,
  title={Benchmark static API call datasets for malware family classification},
  author={Gen{\c{c}}aydin, Buket and Kahya, Ceyda Nur and Demirkiran, Ferhat and D{\"u}zg{\"u}n, Berkant and {\c{C}}ayir, Aykut and Da{\u{g}}, Hasan},
  booktitle={2022 7th International Conference on Computer Science and Engineering (UBMK)},
  pages={1--5},
  year={2022},
  organization={IEEE}
}

@article{cciplak2025fedetect,
  title={FEDetect: A Federated Learning-Based Malware Detection and Classification Using Deep Neural Network Algorithms},
  author={{\c{C}}{\i}plak, Zeki and Y{\i}ld{\i}z, Kaz{\i}m and Alt{\i}nkaya, {\c{S}}ahsene},
  journal={Arabian Journal for Science and Engineering},
  pages={1--28},
  year={2025},
  publisher={Springer}
}

@article{breiman2001random,
  title={Random forests},
  author={Breiman, Leo},
  journal={Machine learning},
  volume={45},
  pages={5--32},
  year={2001},
  publisher={Springer}
}

@inproceedings{chen2016xgboost,
  title={Xgboost: A scalable tree boosting system},
  author={Chen, Tianqi and Guestrin, Carlos},
  booktitle={Proceedings of the 22nd acm sigkdd international conference on knowledge discovery and data mining},
  pages={785--794},
  year={2016}
}

@article{herrera2023dynamic,
  title={Dynamic feature dataset for ransomware detection using machine learning algorithms},
  author={Herrera-Silva, Juan A and Hern{\'a}ndez-{\'A}lvarez, Myriam},
  journal={Sensors},
  volume={23},
  number={3},
  pages={1053},
  year={2023},
  publisher={MDPI}
}

@inproceedings{liu2022towards,
  title={Towards method of horizontal federated learning: A survey},
  author={Liu, Dianqi and Bai, Liang and Yu, Tianyuan and Zhang, Aiming},
  booktitle={2022 8th international conference on big data and information analytics (BigDIA)},
  pages={259--266},
  year={2022},
  organization={IEEE}
}

@article{scikit-learn,
  title={Scikit-learn: Machine Learning in {P}ython},
  author={Pedregosa, F. and Varoquaux, G. and Gramfort, A. and Michel, V.
          and Thirion, B. and Grisel, O. and Blondel, M. and Prettenhofer, P.
          and Weiss, R. and Dubourg, V. and Vanderplas, J. and Passos, A. and
          Cournapeau, D. and Brucher, M. and Perrot, M. and Duchesnay, E.},
  journal={Journal of Machine Learning Research},
  volume={12},
  pages={2825--2830},
  year={2011}
}

@misc{gdpr_eurlex,
  title     = {{General Data Protection Regulation {(GDPR)}}},
  year      = {2020},
  url       = {https://eur-lex.europa.eu/EN/legal-content/summary/general-data-protection-regulation-gdpr.html},
  author    = {{European Union}},
}

@article{mehta2023review,
  title={A review for green energy machine learning and AI services},
  author={Mehta, Yukta and Xu, Rui and Lim, Benjamin and Wu, Jane and Gao, Jerry},
  journal={Energies},
  volume={16},
  number={15},
  pages={5718},
  year={2023},
  publisher={MDPI}
}

@article{lu2024federated,
  title={Federated learning with non-iid data: A survey},
  author={Lu, Zili and Pan, Heng and Dai, Yueyue and Si, Xueming and Zhang, Yan},
  journal={IEEE Internet of Things Journal},
  year={2024},
  publisher={IEEE}
}

@article{zhu2021federated,
  title={Federated learning on non-IID data: A survey},
  author={Zhu, Hangyu and Xu, Jinjin and Liu, Shiqing and Jin, Yaochu},
  journal={Neurocomputing},
  volume={465},
  pages={371--390},
  year={2021},
  publisher={Elsevier}
}

@article{zhao2018federated,
  title={Federated learning with non-iid data},
  author={Zhao, Yue and Li, Meng and Lai, Liangzhen and Suda, Naveen and Civin, Damon and Chandra, Vikas},
  journal={arXiv preprint arXiv:1806.00582},
  year={2018}
}

@article{rey2022federated,
  title={Federated learning for malware detection in IoT devices},
  author={Rey, Valerian and S{\'a}nchez, Pedro Miguel S{\'a}nchez and Celdr{\'a}n, Alberto Huertas and Bovet, G{\'e}r{\^o}me},
  journal={Computer Networks},
  volume={204},
  pages={108693},
  year={2022},
  publisher={Elsevier}
}

@article{fang2023comprehensive,
  title={Comprehensive android malware detection based on federated learning architecture},
  author={Fang, Wenbo and He, Junjiang and Li, Wenshan and Lan, Xiaolong and Chen, Yang and Li, Tao and Huang, Jiwu and Zhang, Linlin},
  journal={IEEE Transactions on Information Forensics and Security},
  volume={18},
  pages={3977--3990},
  year={2023},
  publisher={IEEE}
}

@article{galvez2020less,
  title={Less is more: A privacy-respecting android malware classifier using federated learning},
  author={G{\'a}lvez, Rafa and Moonsamy, Veelasha and Diaz, Claudia},
  journal={arXiv preprint arXiv:2007.08319},
  year={2020}
}

@inproceedings{agrawal2019attention,
  title={Attention in recurrent neural networks for ransomware detection},
  author={Agrawal, Rakshit and Stokes, Jack W and Selvaraj, Karthik and Marinescu, Mady},
  booktitle={ICASSP 2019-2019 IEEE international conference on acoustics, speech and signal processing (ICASSP)},
  pages={3222--3226},
  year={2019},
  organization={IEEE}
}

@online{microsoft2022improving,
  author       = {Microsoft Threat Intelligence},
  title        = {Improving AI-based defenses to disrupt human-operated ransomware},
  year         = {2022},
  month        = {06},
  day          = {21},
  url          = {https://www.microsoft.com/en-us/security/blog/2022/06/21/improving-ai-based-defenses-to-disrupt-human-operated-ransomware/},
  note         = {Microsoft Security Blog}
}

@online{microsoft2017windows,
  author       = {Microsoft Defender Security Research Team},
  title        = {Windows Defender ATP machine learning: Detecting new and unusual breach activity},
  year         = {2017},
  month        = {08},
  day          = {03},
  url          = {https://www.microsoft.com/en-us/security/blog/2017/08/03/windows-defender-atp-machine-learning-detecting-new-and-unusual-breach-activity/?utm_source=chatgpt.com},
  note         = {Microsoft Security Blog}
}

@inproceedings{huang2018tracking,
  title={Tracking ransomware end-to-end},
  author={Huang, Danny Yuxing and Aliapoulios, Maxwell Matthaios and Li, Vector Guo and Invernizzi, Luca and Bursztein, Elie and McRoberts, Kylie and Levin, Jonathan and Levchenko, Kirill and Snoeren, Alex C and McCoy, Damon},
  booktitle={2018 IEEE Symposium on Security and Privacy (SP)},
  pages={618--631},
  year={2018},
  organization={IEEE}
}

@inproceedings{oz2023rob,
  title={$\{$R{\O}B$\}$: Ransomware over modern web browsers},
  author={Oz, Harun and Aris, Ahmet and Acar, Abbas and Tuncay, G{\"u}liz Seray and Babun, Leonardo and Uluagac, Selcuk},
  booktitle={32nd USENIX Security Symposium (USENIX Security 23)},
  pages={7073--7090},
  year={2023}
}

@online{google2021we,
  author       = {Vicente Díaz},
  title        = {We analyzed 80 million ransomware samples – here’s what we learned},
  year         = {2021},
  month        = {10},
  day          = {13},
  url          = {https://blog.google/technology/safety-security/we-analyzed-80-million-ransomware-samples-heres-what-we-learned/?utm_source=chatgpt.com},
  note         = {Google Blog}
}

\end{document}